\documentclass[12pt]{article}
\usepackage{amssymb, amsmath}
\usepackage{hyperref}

\begin{document}
\centerline{\textbf{Exact solutions and their interpretation - session A1}}

\null

\centerline{Ji\v{r}\'{i} Bi\v{c}\'{a}k$^1$ and  Jacek Tafel$^2$} 
\noindent  
{$^1$ Institute of Theoretical Physics,
 Charles University, Prague, Czech Republic, jiri.bicak@mff.cuni.cz}\\ 
{$^2$ Institute of Theoretical Physics, University of Warsaw, 
 Warsaw, Poland, jacek.tafel@fuw.edu.pl}

\null

\begin{abstract}
We report on the oral contributions and give a list of posters presented 
in the session A1 ``Exact solutions and their interpretation'' at the 20th International 
Conference on General Relativity and Gravitation (GR 20) in Warsaw, July 
7-13, 2013. 
\end{abstract}

\vspace{1cm}

\noindent
The workshop A1 included 26 oral contributions, the same number as the workshop ``Numerical relativity and astrophysical applications''. In contrast to most of previous GR-conferences, the largest number of talks (36 and 33) appeared in the parallel sessions  ``Loop quantum gravity...'' and  ``Quantum fields in curved space-time ...''. Sharing the 3-4th ``place'' reveals that the interest in ``exact solutions'' remains strong; a number of young speakers guarantee a bright future. The contribution by Majd Abdelqader was awarded by the Hartle prize for the student presentation. 

There were 5 sessions during first 4 days of the conference. The first two sessions were devoted mostly to various aspects of black-hole and wormhole space-times, then the  session on solutions in higher dimensions and the session on cosmological models, shells and disks followed. The last session involved \textit{miscellanea}, including even some observational aspects. In the following we describe these contributions in more detail.

To reflect the fact that the work on exact solutions has been considerably enhanced during recent decades by the development of software for symbolic calculations, the program was started by the talk        
\textbf{``SageManifolds: a free package for differential geometry and tensor calculus''} by
\textbf{E. Gourgoulhon}. In this work (with M. Bejger),
the development of a new software, SageManifolds \cite{SageManifolds}, for differential geometry and tensor calculus was described. This is a package for 
the modern computer algebra system Sage \cite{Sage}. As Sage, it is free (open source) and is based on the Python programming language. Its distinctive feature is to manipulate the 
various objects at the tensor level, i.e. in a fully coordinate-independent and frame-independent way. This comes along with the capability to deal with various 
charts and vector frames on a given manifold. At the present stage, it is already quite applicable to general relativity, since its functionalities include maps between manifolds, submanifolds, standard tensor calculus (tensor product, contraction,(anti)symmetrization, Lie derivation), Cartan calculus (wedge product and exterior derivative of differential forms), manipulation of affine connections (computation of curvature and torsion) and pseudo-Riemannian metrics (Levi-Civita 
connection, curvature, metric duality and Hodge duality). 

Computers have been a necessary tool also in the contribution by \textbf{M. Abdelqader ``Analyzing and visualizing the Kerr vacuum via gradient flows''} (a joint work with K. Lake). 
The authors  explored the global structure of the Kerr spacetime using a new visualization and analysis tool based on
gradient fields of scalar invariants. They observed that a structure of the Kerr vacuum  outside the horizon does not vary smoothly with the spin parameter, but goes through a significant
qualitative change at some specific ``transitional'' values of the spin parameter. The number of the gradient fields' critical points and their index, or winding number, along the axis of symmetry changes at these values. These gradient fields represent the cumulative tidal and frame-dragging effects in an
observer independent way. The results have possible applications in theoretical astrophysics (alternative methods to extract mass and angular momentum in numerical relativity, understanding effects of different spin parameter values etc).

The classical model of electron with over-extreme Kerr-Newman metric outside was presented in the talk \textbf{``Regularized Kerr-Newman solution as a model spinning particle''} by \textbf{A. Burinskii}. The naked Kerr singular ring is replaced by a regular material source consistent with external KN geometry. L\'opez (1984) determined the boundary of the KN source, $r=r_e$, such that the external KN metric may be continuously matched with a flat interior for $r<r_e.$  The boundary $r_e = e^2/2m $ determines a ``bubble'' with  a special distribution of charged matter. This model was developed in \cite{BurSol} as a gravitating soliton model, in which the thin shell model is replaced by a domain wall which provides a phase transition from external KN fields to an internal false-vacuum state. The  superconducting Higgs-like field inside the bubble expels the electromagnetic field. The model exhibits some properties of the extended dressed electron, as well as Zitterbewegung of the point-like naked electron.

\textbf{``Scalar multi-wormholes''} were presented by  \textbf{S. V. Sushkov} -- joint work with  A. I. Egorov and P. E. Kashargin. 
The authors  constructed exact axially symmetric solutions describing $N$ wormholes  with the phantom scalar field (the kinetic energy term enters action with the ``opposite sign''). 
Space-time corresponding to these solutions has $2^N$ asymptotically flat regions connected by $N$ throats. Unlike in a linear superposition of $N$ collinear Schwarzschild black holes one needs no singular struts to provide a static character of the configuration. The space-time is regular, with no  event horizons. This is possible because of exotic properties of the phantom scalar field violating standard energy conditions.

``Unorthodox'' wormhole configurations were also discussed in the talk  \textbf{``Off--Diagonal Wormhole Deformations in Modified Massive Gravity''} by  \textbf{S. I. Vacaru}. The author studied
deformations of ``prime'' metrics in the Einstein theory (e.g. wormhole configurations) into exact solutions in $f(R,T)$--modified and massive/bi-metric gravity theories. The anholonomic frame deformation method was applied to construct solutions in modified gravity theories (MGT) (see, e.g.  \cite{v1,v3}), when the field equations decouple in a general form. This allows  to construct generic off--diagonal solutions  depending on all spacetime coordinates via generating and integration functions and constant parameters. For certain well defined conditions on generating functions and non--integrable constraints, the MGTs effects can be encoded as (effective) Einstein manifolds. Using nonholonomic deformations with  ellipsoid/toroid and/or solitonic symmetries, one can generate wormhole like objects matching external black ellipsoid-de Sitter geometries.

The contribution \textbf{``Static spherical black holes with scalar field''}  of \textbf{J. Tafel} 
refers to static spherically symmetric solutions of the Einstein equations with   minimally coupled scalar field $\varphi$. Asymptotically flat or anti-de Sitter black holes and particle like solutions are considered \cite{t}. A potential  $V(\varphi)$ is not a'priori prescribed. It is shown that the Einstein equations  define  metric, $\varphi$ and $V$  in terms of a free function $\rho$ which is monotonically growing, convex, has a zero point and behaves like  $r-3M$ when  $r$ is large.  Given $\rho$  a dependence of  $V$ on $\varphi$ is given in a parametric way. 
This description allows to generalize  no-hair theorems (Bekenstein 1972 and later developments)  to the AdS asymptotic and to obtain new properties of solutions for both considered asymptotics. In particular, for any potential $V(\varphi)$ the total ADM mass $M$ has to be positive if $\varphi$ is nontrivial.  In the case of black holes a radius $r_h$ of the Killing horizon has to obey the Penrose inequality $r_h\leq 2M$  and the surface gravity is nonzero. In generic case $\varphi$ is finite at singularity and $V$ is infinite. Regular potentials $V(\varphi)$ are admitted, bounded or unbounded, if   derivatives of $\rho$ at $r=0$ (singularity) are properly chosen.  Examples of solutions were presented in a form of plots.

In the talk \textbf{ ``Relativistic charged spheres: Exact solutions for stars, regular black holes and quasi-black holes''}, \textbf{J. P. S. Lemos} presented results obtained with V. T. Zanchin on a class of exact solutions of spherically symmetric distributions of charged matter matched smoothly to a
Reissner-N\"{o}rdstrom solution. The interior parameters are
given in terms of the radius of the configuration 
and the exterior global ADM mass and electric
charge. The solutions were found by Guilfoyle (Gen. Rel. Grav. 31, 1645). Some of their properties  were analyzed in \cite{lemoszanchin2010}. Now the authors found  further properties and shown that
they contain a rich variety of configurations and
spacetimes: undercharged, extremally  charged and
overcharged stars, regular black holes, and quasi-black holes. These latter may be considered as frozen stars, i.e., collapsed stars with their boundary surfaces looming at their
own gravitational radii. The spectrum of solutions also contains ``bridge spacetimes'' with a negative electromagnetic field energy of the type considered by Einstein and Rosen (as a non-singular ``electron''). The solutions yield a generalization of the
Buchdahl limit in the uncharged case.

In the talk \textbf{ ``The conformal cousin of the Husain-Martinez-Nu\~nez spacetime'', V. Faraoni} discussed
￼￼￼ exact solutions of gravitational field equations which can be interpreted as black holes embedded in cosmological ``backgrounds''. Only few such solutions are known.  Although they are probably pathological to some extent, they can tell us something about the theory which would be otherwise hard to guess. It is disappointing that  the known solutions are usually reported, but not studied further nor interpreted physically. For example, it is not known whether these dynamical inhomogeneities represent black holes (as defined by the existence of apparent horizons) or naked singularities.  There are reasons to study these solutions in modified gravity theories. In \cite{FaraoniZambrano12},  a 2-parameter, spherically symmetric, inhomogeneous cosmology in Brans-Dicke theory, 
obtained by conformally transforming the Husain-Martinez-Nu\~nez (massless, minimally coupled) scalar field solution of the Einstein equations, is studied. Surprisingly, depending on values of the parameters, the solution describes a black hole dressed by apparent horizons which appear or disappear in pairs, or a wormhole, or a naked singularity.  The reason why there is not a one-to-one correspondence between conformal copies of this metric was discussed.

Results presented by \textbf{N. G\"urlebeck} in the talk \textbf{ ``Tidally Distorted Black Holes''} concern a generalization of 
a well-known no-hair theorem to more astrophysical
settings. In the static case, the standard version of this theorem states that an uncharged and isolated black hole is fully characterized by its mass. The generalization includes black holes, which are distorted due to the gravitational field of  additional sources. As a starting point served  metric that describes all static, axially symmetric and distorted black holes near the horizon (Geroch and Hartle 1982). Using this metric, G\"urlebeck proved that no multipole moment is induced in the black holes by the external sources in spite of the distortion. Such deformed black holes contribute to the asymptotic multipole moments only by a mass monopole. The key tool in the proof are source integrals of the asymptotic multipoles \cite{Gurlebeck_2012_Gurlebeck}. This approach does not
only lead to the generalized version of the no-hair theorem but it also solves a debate on the Love numbers of black holes. They measure the distortion of a star; however, they vanish in the limit of a black hole. The aforementioned proof shows that this result holds in full general relativity, it is not due to  approximations taken in previous works.

In the talk of \textbf{G. A. Alekseev} \textbf{``Dynamics of black holes in AdS2 x S2 spacetimes (Bertotti-Robinson universes)''}
 new applications of his  monodromy transform approach and corresponding linear singular integral equation method for solving of Einstein-Maxwell equations (see \cite{Alekseev:2011} and the references there) were presented. Two new exact solutions of Einstein-Maxwell equations were found in a very simple explicit form. One of these solutions generalizes the solution found earlier in \cite{Alekseev-Garcia:1996} for a Schwarzschild black hole in a static position in  (asymptotically) Bertotti-Robinson pure magnetic universe. It describes a Schwarzschild black hole in a ``geodesic'' motion along the magnetic field in the gravitational field of this universe. Another solution describes a Reissner-N\"{o}rdstrom black hole accelerated by the electric field of (asymptotically) Bertotti-Robinson pure electric universe. In both cases, the condition of the absence of conical singularities on the axis of symmetry is satisfied. This determines the ``geodesic''  character of motion of a neutral black hole along the magnetic field and acceleration of a charged black hole in the external asymptotically homogeneous electric field. The acceleration of a black hole is related to  its charge and strength of the external electric field.

The talk \textbf{``Black holes in 4-dimensional supergravity''} of \textbf{D. Chow} was devoted to
 new black hole solutions in four-dimensional supergravity \cite{ChowCompere}.  They are solutions of the so-called ``STU'' model of $\mathcal{N} = 2$ supergravity coupled to 3 vector multiplets.  Upon reduction by means of a timelike symmetry to 3 dimensions and dualizing vectors to scalars, the theory is three-dimensional gravity coupled to an $SO(4, 4)$ coset model.  Acting with $SO(4, 4)$ generators and then lifting back to 4 dimensions generates  new four-dimensional solutions.  This generalizes the method of ``Harrison transformations'' for the Einstein--Maxwell theory.  If one starts with the Kerr--Taub--NUT solution and uses an appropriate set of generators the Kerr--Taub--NUT solution with 4 electric charges and 4 magnetic charges is generated.  Setting the NUT charge to zero  gives asymptotically flat black holes that generalize the Kerr--Newman solution.  This provides a seed solution to generate the general asymptotically flat, rotating black hole of $\mathcal{N} = 8$ supergravity with 28 electric charges and 28 magnetic charges.  There are also some generalizations to $\textrm{U}(1)^4$ gauged supergravity that are asymptotically anti-de Sitter.  These include new examples of static anti-de Sitter black holes with 4 electric charges and 4 magnetic charges and with spherical or planar horizons.

In the talk \textbf{ ``Black rings in higher dimensions'', J. Kunz}
discussed a new type of
black holes in higher dimensions: black rings. Their existence was first suggested by Myers and Perry (MP).  
Emparan and Reall then constructed rotating 5-dimensional
vacuum black rings, balancing the attractive forces of gravity
and string tension by the centrifugal force.
The rings possess two branches,
a branch of thin black rings and a branch of fat black rings.
The common phase diagram of MP black holes
and black rings, both with a single angular momentum, reveals that uniqueness is violated for these vacuum black objects.
In a small region of the phase diagram
there reside three different solutions: one black hole and two black rings, all three of them have the same mass and the same angular momentum. Recently, an analogous phase diagram was obtained for 6-dimensional black objects \cite{Kleihaus:2012xh}.
Again two branches of black rings exist. 
Here the branch of fat black rings does not reach the
MP solutions in a singular configuration.
When the inner radius of the ring shrinks to zero
a singular horizon topology changing configuration is approached.
This critical configuration cannot be reached numerically.
At the horizon topology changing configuration one expects
a merger with the first still hypothetical branch of pinched black holes.
These branches should arise because of the Gregory-Laflamme type instabilities of
the singly rotating MP solutions in 6 and more dimensions.

Higher dimensional black holes were also discussed in the talk
\textbf{``Maximal slices of five dimensional black holes''} of \textbf{H. K. Kunduri}. In four dimensions
initial data for Einstein's equations are specified by a spacelike hypersurface $\Sigma$ with metric tensor $h$ and the second fundamental form $K$. For globally hyperbolic $M$, one can write $M \backsimeq \mathbb{R} \times \Sigma$ where $\Sigma$ is a Cauchy slice.  The topological censorship (J. Friedman et al, PRL 71, 1486) asserts that any non-simply connected topological structures in $\Sigma$ must be hidden behind horizons. For example, a maximal slice of the Kerr black hole has topology $\Sigma \backsimeq \mathbb{R} \times S^2$. In five dimensions, however, topological censorship is far less restrictive. Stationary black holes admitting a rotational isometry  may have horizon topology $H \backsimeq S^3$ (and quotients), $S^1 \times S^2$, and connected sums thereof.  In the recent work \cite{GR20_A1_KunduriRef3} the author and collaborators  investigated the topology and geometry of maximal slices of the Myers-Perry black hole, the vacuum black ring and the black Saturn solution  and demonstrated that $\Sigma$ is diffeomorphic to $\mathbb{R} \times S^3, S^2 \times D^2 - \{pt\}$  and $S^2 \times D^2 \#\mathbb{R}^4\#B^4$, respectively.  Despite being simply connected, the latter two examples have interesting topological structures such as non-trivial 2-cycles. As much of this analysis is at the topological level, this work was also extended to  black hole configurations for which explicit geometries are not yet known.

An explicit \textbf{``Algebraic classification of Kundt geometries in four and higher dimensions''} for which there exists a privileged non-expanding multiple ``Weyl aligned null direction'' (WAND) was presented by \textbf{J.~Podolsk\'y} -- a joint work with R. \v{S}varc. No field equations were used, the results apply  to any metric theory of gravity which admits the Kundt spacetimes.  All Kundt geometries are of type I(b). Simple necessary and sufficient conditions under which WAND becomes double, triple or quadruple  were derived. All possible algebraically special types, including the refinement to subtypes, were identified. To illustrate the classification scheme, it was applied to some subfamilies of the Kundt class, e.g. pp-waves, the VSI space-times and a generalization of the Bertotti--Robinson, Nariai, and Plebanski--Hacyan space-times in any dimension (see \cite{PodolskySvarcA,PodolskySvarcB}).

In four dimensions, a fundamental connection between geometric optics and the algebraic structure of the Weyl tensor is provided by the Goldberg-Sachs (GS) theorem, which relates algebraically special spacetimes to shearfree geodesic null congruences. \textbf{``An extension of the Goldberg-Sachs theorem in five and higher dimensions''} was discussed by \textbf{M. Ortaggio}.
The standard formulation does not admit a straightforward extension to higher dimensions, however, results on possible generalizations have been obtained. In particular, in this talk constraints on optical properties of null congruences ``multiply aligned'' with the Weyl tensor, which extend in $D>4$ the ``shearfree part'' of the GS theorem were summarized \cite{Ortaggioetal12,OrtPraPra12}. Possible canonical forms of the ``optical matrix'' were presented explicitly in five dimensions (or in six dimensions for twist-free null congruences).

\textbf{``Higher dimensional gravitating fluids''} were studied by \textbf{Y. Nyonyi}, in collaboration with  S. Maharaj and K. Govinder. They considered shear-free spherically symmetric cosmological models with heat flow and charge defined on
an (N + 2)-dimensional manifold. The pressure isotropy condition is a nonlinear PDE which was treated  using the Lie group theoretical approach. Symmetry generators that leave the equation invariant were found. Their knowledge allows to obtain  exact solutions for the gravitational potentials. They contain  earlier solutions without charge.

In general relativity there has been considerable interest in the different types of singularities that the Friedmann-Robertson-Walker (FRW) space-times admit,
namely Big Bangs, Big Crunches, Big Rips, Sudden Singularities and Extremality Events. The scale factor, and its
derivatives, are pivotal in determining the nature of singularities in the FRW space-times. Following Catto\"en and Visser (Class. Quantum Grav. 22, 4913) \textbf{S. M. Scott}, in the joint contribution with P. Threlfall \textbf{``The conformal structure of past and future end states of FRW space-times''}, considered FRW space-times with a scale factor expressed as a generalized power series. Inspired by the definition of an Isotropic Past Singularity by Goode and Wainwright for an \textit{initial} singularity with regular conformal structure (an appropriate beginning for the Quiescent Cosmology), H\"ohn and Scott \cite{Scott2009} defined generic conformal singularities for possible isotropic and anisotropic cosmological \textit{future end states}. The singularities found by Catto\"en and Visser for the FRW space-times were analyzed and compared with those models admitting the generic conformal singularities of an Isotropic Future Singularity or a Future Isotropic Universe.

\textbf{``An exact smooth Gowdy-symmetric generalized Taub-NUT solution''}, presented by \textbf{J. Hennig}, belongs to a class of inhomogeneous cosmological models with spatial $S^3$-topology \cite{BeyerHennig2012}. They have the Cauchy horizon at $t=0$ and, in general, develop the second Cauchy horizon at $t=\pi$. For particular choices of data at $t=0$ a singularity forms at $t=\pi$. The author found a three-parameter family of the solutions which contains the two-parametric Taub solution. To this end Sibgatullin's integral method  (Oscillations and Waves, 1984) was used to solve an initial value problem for the hyperbolic Ernst equation. For a special choice of the parameters, the solution contains a curvature singularity with directional behaviour. For other parameter choices, the maximal globally hyperbolic region is singularity-free. Furthermore, several extensions of the solution through the Cauchy horizons have been constructed \cite{BeyerHennig2013})

\textbf{``New exact solutions for spherical objects with charge 
and anisotropic pressures''}, representing generalizations of the Tolman VII solution, were presented by \textbf{A. Raghoonundun}, in collaboration with D. Hobill. These solutions have appropriate physical properties. The approach provides  means for computing a zero
temperature equation of state relating pressure and density in the
star. For a wide range of accessible parameter values, these models manage to predict values associated with observed neutron stars in accordance with  causality and energy conditions.  The anisotropic pressure solution labeled TVIIa and the charged version labeled TVIIac are completely defined by their metric functions. The mass density depends quadratically on the radial coordinate. If the anisotropic pressures vanish TVIIa reduces to the Tolman VII solution and TVIIac reduces to the solution of Kyle and Martin (N. Cim. A 50, 583).

An infinite family of new solutions of the Einstein-Maxwell equations representing \textbf{``Magnetized axially symmetric thin dust disks in conformastatic spacetimes''} was presented by  \textbf{G. Gonzalez}. The disks are made of material
sources with a surface conduction current. The solutions are obtained by expressing the metric function
and the magnetic potential in terms of an auxiliary function satisfying the Laplace equation. The surface energy-momentum and the surface current density of the disk are obtained by using the
formalism of tensorial distributions. A simple particular model was presented in which the energy is well behaved everywhere and the energy-momentum tensor satisfies all standard energy conditions. Although the disks are  infinite, their total mass is
finite.

Thin shells, rather than thin disks, were discussed in the talk of \textbf{M.A. Ramirez} entitled \textbf{``Splitting thin shells and the evolution of distributional solutions of Einstein's
equations''}. A number of solutions in the sense of distributions involving thin shells was presented and their stability against separation of their constituents was analyzed. First the spherically symmetric shells made of Vlasov matter were considered and the stability analysis against fragmentation done. Next, it was shown that dynamical shells may be composed of particles orbiting at different angular velocities, but in order to evolve stably as a single shell the angular momentum distribution cannot be arbitrary \cite{GleRam09}. There are also {\texttt splitting} solutions, in which the original shell smoothly splits into a number of emergent shells. For a given initial data set, both the original shell without splitting and the splitting solution solve the Einstein equations coupled to matter, which illustrates a lack of uniqueness for the Cauchy problem. The unstable non-splitting solutions appear not physical as they may not be thin shell limits of thick shell solutions \cite{GleRam09} \cite{GleRam10} (for generalizations, see Ramirez M A, arXiv 1207.6810[gr-qc]).

Test particles follow geodesic trajectories in space-time. This is the leading order result from a multipolar expansion of the energy-momentum tensor of a test body. At next order, including pole and dipole terms, the Mathisson-Papapetrou (MP) equations of motion for spinning test particles play the role. Also, the Spin Supplementary Condition (SSC) must be specified to fix the worldline of the body. In the contribution \textbf{``Spinning test particle trajectories in de Sitter space-time''} (joint work  with J. Gair and S. Babak) \textbf{R. Cole } presented an analytic solution to the MP equations in de Sitter spacetime using the Frenkel-Pirani SSC and the w-condition. He also discussed methods of studying the motion of spinning test particles in black hole spacetimes,
using the de Sitter solution. This has applications in the generation of waveform templates for extreme-mass-ratio inspirals since the presence of spin in the compact object can lead to a significant dephasing over the lifetime of the inspiral.

The subject of gravitational lensing has developed to a very high degree of sophistication, nevertheless, it has focused on lensing in stationary systems. In his talk \textbf{``Gravitational lensing by gravitational waves'', A. Harte} discussed lensing by gravitational waves. Assuming the planar symmetry, various results can be computed exactly. Images of objects are typically distorted, change brightness and color, and appear to move across an observer's sky due to the waves. Multiple images can also be formed. Even in the simplest cases involving short wave packets, dramatic effects occur generically. Essentially any observer eventually reaches a point in time at which light from the entire universe is momentarily concentrated to a single point in the sky. Formally, this light is infinitely blueshifted and was emitted in the infinitely distant past. Later, the contributions to this flash from different sources appear to separate from each other, dim, and then redshift away.

In the talk of \textbf{O. Yu. Tsupko}, based on the work with G. S. Bisnovatyi-Kogan, \textbf{``Gravitational deflection of light ray in plasma''} was studied. In absence of a refraction, in a homogeneous dispersive medium, the gravitational deflection is qualitatively different from the vacuum case and the deflection angle depends on the photon frequency. Simple analytical formula for the gravitational light deflection in the Schwarzschild metric, in presence of a homogeneous plasma, was obtained \cite{BKTs2010}.  The effect is different from the vacuum case but it is significant only for the radio waves. Using these results, a model of gravitational lensing in plasma was developed. In the example of the Schwarzschild point-mass lens, instead of two point-like images with complicated spectra, there should be two line images, formed by  photons with different frequencies, which are deflected by different angles. A more general approach, considering an inhomogeneous plasma, was also developed \cite{BKTs2010}, \cite{BKTs2013}.

\noindent

In classical GR there are many solutions describing naked singularities and static or stationary configurations with nom vanishing energy momentum. \textbf{D. Malafarina} analyzed \textbf{``Observational features of perfect fluid sources with a singularity at the center''}. These solutions can be obtained from gravitational collapse under  simple assumptions
\cite{2}. It is interesting what observational features  distinguish them from black holes of the same mass.
In this contribution compact objects (with interiors described by the Tolman perfect fluid solutions), surrounded by accretion disks, were considered. The luminosity spectrum as received by distant observers exhibits a tail at high frequencies that is absent in the case of black holes \cite{2}. Furthermore, simulations of the $K_\alpha$ iron line of absorption show very different behavior in the two cases \cite{3}.

In the last oral presentation in the workshop \textbf{M. O. Katanaev} asked what is \textbf{``Point massive particle in General Relativity''}. His recent work \cite{Katana13}  indicates that the Schwarzschild solution in isotropic coordinates can be considered as the asymptotically flat  solution of Einstein's equations with $\delta$-type
energy-momentum tensor corresponding to a point particle. 
The solution is understood as a distribution. Metric components are locally integrable functions. The gravitational attraction
at large distances is replaced by repulsion at the particle neighborhood.
The author claims that if the requirement of geodesic completeness is changed into the requirement of geodesic completeness only in the physical sector then the space-time needs no continuation \cite{Katana13B}.

\bigskip

In addition to talks 10 \textbf{posters} were exhibited in the framework of the session A1. We list them below with  indication of persons presenting them. Description of the posters can be found on the conference page (gr20-amaldi10.edu.pl).
\begin{enumerate}
\item 
Gonzalez, G., \textbf{Pimentel, O.}: 
Axially symmetric relativistic thin disks immersed in spheroidal matter
haloes
\item 
Gonzalez, G., \textbf{Navarro, A.}: 
Relativistic static thin disks with electrically and magnetically polarized
material source
\item 
\textbf{Sunzu, J.}: 
Anisotropic charged exact models
\item 
\textbf{Abebe, G.}, Govinder, K., Maharaj, S.: 
Lie symmetries for a radiating star in conformally flat spacetime
\item 
\textbf{Chilambwe, B.}, Hansraj, S., Maharaj, S.: 
Exact interior solutions in Einstein-Gauss-Bonnet gravity
\item 
Frolov, B.,\textbf{ Febres, E.}: 
Spherically symmetric solution of gravitation theory with Deser-Dirac scalar
field in Riemann-Weyl space
\item 
\textbf{Shcherban, V. N.}:
Investigation of Plane Waves in Torsion Poincare Gauge
Theory of Gravity
 
\item 
\textbf{Pravda, V.}:  
On the Goldberg-Sachs theorem in five dimensions

\item 
Korkina, M., \textbf{Iegurnov, O.}: 
Matching of the de Sitter solution and solution for perfect fluid with
nonuniform pressure

\item 
\textbf{Bradley, M.}, Machado Ramos, M.:                                                                    
Killing vector analysis in GHP formalism of conformally flat pure radiation
metrics with negative cosmological constant
\end{enumerate}

\section*{Acknowledgments.}
J.B. acknowledges partial support from the Grant GA\v{C}R No. 14-37086G
from the Czech Science Foundation, which enabled him to visit GR20, and the
hospitality
of the Hebrew University, Jerusalem, in particular of Tsvi Piran and Joseph
Katz, where a part
of this report was written.


\begin{thebibliography}{99}

\bibitem{SageManifolds}\url{http://sagemanifolds.obspm.fr/}

\bibitem{Sage}\url{http://sagemath.org/}

\bibitem{BurSol}  Burinskii, A.: Regularized Kerr-Newman Solution as a Gravitating Soliton,
 J. Phys. A \textbf{43 }, 392001 (2010), arXiv: 1003.2928 [gr-qc]

\bibitem{BurQ}  Burinskii, A.: Gravity vs. Quantum theory: Is electron really pointlike?, J.  Phys.: Conference Series, \textbf{343}, 012019 (2012), arxiv:1112.0225 [gr-qc]

\bibitem{v1} Vacaru, S., Singleton, D.: Warped, anisotropic wormhole/soliton configurations in vacuum 5D gravity, Class. Quantum Grav. \textbf{19},  2793 (2002) 

\bibitem{v3}
Vacaru, S.: Hidden symmetries for ellipsoid-solitonic deformations of Kerr-Sen black holes and quantum anomalies, Eur. Phys. J.  \textbf{C 73}, 2287 (2013)
\bibitem{t}
Tafel, J.: Static spherically symmetric black holes with scalar field, to appear in Gen. Rel. Grav., arXiv:1112.2687 [gr-qc]

\bibitem{lemoszanchin2010}  Lemos, J.P.S., Zanchin, V.T.: Quasiblack
holes with pressure: Relativistic charged spheres as the frozen
stars, Phys. Rev. D \textbf{81}, 124016 (2010), arXiv:1004.3574 [gr-qc]

\bibitem{FaraoniZambrano12} 
 Faraoni, V.,  Zambrano Moreno, A.F.: Interpreting the conformal cousin of the Husain-Martinez-Nu\~nez solution, Phys. Rev. D \textbf{86}, 084044 (2012)

\bibitem{Gurlebeck_2012_Gurlebeck} G\"urlebeck, N.: Source integrals for
multipole moments in static and axially symmetric spacetimes, submitted to Phys. Rev. D,
arXiv:1207.4500 [gr-qc]


\bibitem{Alekseev:2011} Alekseev, G.A.: Thirty years of studies of integrable reductions of Einstein's field equations, Proc.  XII Marcel Grossmann Meeting, World Scientific, Singapore (2011), arXiv:1011.3846v1 [gr-qc]

\bibitem{Alekseev-Garcia:1996} Alekseev, G.A., Garcia, A.A.: Schwarzschild black hole immersed in a homogeneous electromagnetic field, Phys. Rev. {\bf D53},  1853-1867 (1996)

\bibitem{ChowCompere} 
  Chow, D.D.K., Compère, G.:
  Seed for general rotating non-extremal black holes of $\mathcal{N}=8$ supergravity,
  arXiv:1310.1925 [hep-th].
  
\bibitem{Kleihaus:2012xh} 
  Kleihaus, B., Kunz, J., Radu, E.:  Black rings in six dimensions,
  Phys.\ Lett.\ B {\bf 718}, 1073 (2013)


\bibitem{GR20_A1_KunduriRef3}
  Alaee, A., Kunduri, H.K., Mart\'{\i}nez-Pedroza, E.:
  Notes on maximal slices of five-dimensional black holes,
  arXiv:1309.2613 [gr-qc]

\bibitem{PodolskySvarcA} Podolsk\'y, J., \v Svarc, R.:
Explicit algebraic classification of Kundt geometries in any dimension,
 Class. Quantum Grav. {\bf 30} 125007 (2013)

\bibitem{PodolskySvarcB}  Podolsk\'y, J., \v Svarc, R.:
Physical interpretation of Kundt spacetimes using geodesic deviation,
Class. Quantum Grav. {\bf 30} 205016 (2013)


\bibitem{Ortaggioetal12}
Ortaggio, M., Pravda, V., Pravdov\'a, A.,  Reall. H.S.:
 On a five-dimensional version of the {G}oldberg-{S}achs theorem, Class. Quantum Grav. \textbf{29}, 205002 (2012)

\bibitem{OrtPraPra12}
Ortaggio, M., Pravda, V., Pravdov\'a, A.:  On the {G}oldberg-{S}achs in higher dimensions in the non-twisting
  case, Class. Quantum Grav. \textbf{30}, 075016 (2013)


\bibitem{Scott2009}  H\"ohn, P.A., Scott, S.M.: Encoding cosmological futures with conformal structures, Class. Quantum Grav. \textbf{26}, 035019 (2009) 

 \bibitem{BeyerHennig2012}
 Beyer, F., Hennig, J.: Smooth Gowdy-symmetric generalized Taub-NUT solutions, 
 Class. Quantum Grav. {\bf 29}, 245017 (2012)

 \bibitem{BeyerHennig2013}
 Beyer, F., Hennig, J.:
 An exact smooth Gowdy-symmetric generalized Taub-NUT solution, 
 submitted (2013)

 


\bibitem{GleRam09} Gleiser, R.J., Ramirez, M.A.: On the dynamics of thin shells of counter-rotating particles, Class. Quantum Grav.  \textbf{26}, 045006 (2009)
\bibitem{GleRam10} Gleiser, R.J., Ramirez, M. A.: Static spherically symmetric Einstein-Vlasov shells made of particles with a discrete set of values of their angular momentum, Class. Quantum Grav. \textbf{27}, 065008 (2010)



\bibitem{BKTs2010}
Bisnovatyi-Kogan, G.S., Tsupko, O.Yu.: Gravitational lensing in a non-uniform plasma, Monthly Notices of the Royal Astronomical Society  \textbf{404}, 1790 (2010)

\bibitem{BKTs2013}
Tsupko, O.Yu., Bisnovatyi-Kogan, G.S.: Gravitational lensing in plasma: Relativistic images at homogeneous plasma, Phys. Rev. D \textbf{87}, 124009 (2013)


\bibitem{2} Joshi, P.S., Malafarina, D., Narayan, R.: Distinguishing black holes from naked singularities through their accretion disk properties, 
arXiv:1304.7331 [gr-qc]

\bibitem{3} Bambi, C., Malafarina, D.:
$K_\alpha$ iron line profile from accretion disks around regular and singular exotic compact objects,
Phys.  Rev. D {\bf 88}, 064022 (2013), arXiv:1307.2106 [gr-qc]


\bibitem{Katana13}
Katanaev, M.O.:
 Point massive particle in general relativity, Gen. Rel. Grav. \textbf{45}, 1861--1875 (2013)

\bibitem{Katana13B}
Katanaev, M.O.:
Passing the Einstein--Rosen bridge, arXiv:1207.3481 (2012).
\end{thebibliography}
\end{document}